\documentclass[12pt]{article}%
\usepackage{amssymb}
\usepackage{amsmath}%
\usepackage{amsfonts}%
\usepackage{graphicx}
\providecommand{\U}[1]{\protect\rule{.1in}{.1in}}

\begin{document}
 \bigskip\begin{titlepage}
\begin{flushright}
PUPT \\
hep-th/yymmnnn
\end{flushright}
\vspace{7 mm}
\begin{center}
\huge{Quantum instability of the de Sitter space}
\end{center}
\vspace{10 mm}
\begin{center}
{\large
A.M.~Polyakov\\
}
\vspace{3mm}
Joseph Henry Laboratories\\
Princeton University\\
Princeton, New Jersey 08544
\end{center}
\vspace{7mm}
\begin{center}
{\large Abstract}
\end{center}
\noindent
We continue to investigate various instabilities of the fixed backgrounds related to
the de Sitter space. It is shown that in many cases  the in/in perturbation theory contains
IR/UV mixing and thus is non-renormalizible. The application of this result to the global de Sitter
space leads to the conclusion that even massive particles generate IR divergence and the huge back
reaction. The expanding universe is also unstable but in a weaker sense. We further discuss ,the
strange features of the Gibbons-Hawking radiation and its relation to the above instabilities. .
\begin{flushleft}
September 2012
\end{flushleft}
\end{titlepage}\bigskip\ 

\section{Introduction}

In the naive approach to quantum gravity vacuum fluctuations generate a
cosmological constant, the size of which is determined by the ultraviolet
cut-off, the Planck length. At the same time the cosmic acceleration indicates
that the observable cosmological constant is determined by the size of the
universe , which can be viewed as an ultimate infrared cut-off. The problem of
the cosmological constant can be formulated as following - \emph{why }%
$\Lambda$\emph{\ is determined by the IR and not by the UV scale ? }This gives
a strong hint that the resolution of the cosmological constant problem must be
based on the relevant infrared interactions.

There are many examples in quantum field theory ( say the Yang-Mills theory)
in which such interactions deform the classical equations of motion beyond
recognition. The cosmological term in the Einstein equation may experience the
same fate, since it doesn't contain derivatives of the metric and thus may be
relevant in the infrared \cite{pol81}.

Unfortunately, we are still lacking the formal apparatus , needed for this
problem. In particular, the scenario in \cite{pol81} is based on the domains
where the vacuum expectation value of the metric $<g_{\mu\nu}>=0.$ It is not
clear how to treat it quantitatively and the problem remains unsolved.
However, it is worth mentioning, that the common counter-argument against this
approach, namely that the scale factor of the metric can be removed by a gauge
transformation and thus is not physical, is incorrect. \ In the zero metric
domain the scale factor can't be removed , just like the unitary gauge in the
standard model can't be used in the unbroken phase where the expectation value
of the Higgs field is zero.

Another approach to the problem (taken in this article) is to study possible
instabilities of the backgrounds with positive curvature. Such backgrounds are
in many respects similar to the simpler case of the constant electric fields
in the flat space ( while the negative curvature case is analogous to magnetic
field ). It is very well known that the electric field makes the empty space
unstable by producing pairs of particles. These pairs eventually screen the
source of the field and the production stops.The energy needed for the pair
production is taken from the electric field which becomes oscillating and decreasing.

If we believe the analogy, we should expect a depletion of the original
curvature, due to the back reaction of the particles it has generated. More
precisely, the constant curvature means that the space is either expanding or
contracting with constant acceleration. As we will see , there is a constant
particle creation in this case . The gravitational attraction between the
particles will slow down the acceleration , producing something like radiation damping.

The most common objection to this picture is that the de Sitter space seems
stable. Indeed , consider a theory of a massive scalar interacting field
$\varphi(x)$ in the dS background. The back reaction is defined by the one
point functions , like the energy- momentum tensor or simply by the operators
like $\langle\varphi^{2}(x)\rangle$ . However , the isometries of the de
Sitter space imply that such objects do not depend on $x$ \ and thus the back
reaction never get large. In contrast, the constant electric field induces the
current which grows linearly with time , since the production goes with the
constant rate. Let us notice, however, that even in the electric case this
blow-up of the current is in the apparent contradiction with the time-
translation invariance of the constant electric field.

One can also argue that the field theory can be defined on the Euclidean
version of dS ( a simple sphere ) and then analytically continued to the
Lorentzian regime by the Wick rotation. Once again it results, at least
perturbatively, in stability, since the theory on a sphere is clearly well defined.

As we will see below, both arguments are incorrect. First of all, as was
argued in \cite{pol10} \ , there is an obstruction to the Wick rotation.
Second , we will see that the isometries of the dS being unstable are
spontaneously broken in the above case. This fact allows the back reaction to
become large.

This situation is very similar to the one we have in the case of the
Schwarzcshild black hole. In this case the Euclidean field theory is also
well-defined in all orders of perturbation theory. However, it misses the
black hole evaporation. \ Similarly, the field theory on a sphere misses the "
evaporation" of the positive curvature of the dS space.

Let us mention that the " electric " analogue of the black hole are
supercharged nuclei. In this case it decays by creating bounded electrons and
run away positrons, until it gets neutralized. Here again the analog of the dS
symmetry is time- translation invariance in the presence of time independent
electric field. It is spontaneously broken and the back - reaction current
becomes proportional to time, thus getting large.

To some extent these questions have been discussed in \cite{pol11} , but many
points remained unclear. In this paper we will concentrate on the origin of
the infrared divergences, their unexpected and unusual mixture with the
ultraviolet, and on the peculiar mechanism which destroys isometries. As
before, we will warm up with the electric fields and then proceed analyzing
Poincare patches and finally global de Sitter space. The latter case is
especially interesting, since while being classically stable \cite{anderson}
its isometries are destroyed by the IR/UV mixing , mentioned above.

We will stress that the new general phenomenon, responsible for the large back
reaction in unstable vacua is the non- commutativity of the UV and IR limits.
As it often happens, non-commutative limits are the source of many confusions.

\bigskip There are large number of different approaches to these problems in
the literature and it is sometimes hard to establish connection with the
results of this paper. Some, very incomplete, list of references can be found
in \cite{pol11}.

\section{Electric IR/UV mixing}

In this section we will discuss instabilities in the strong electric fields.
There are hundreds if not thousands of papers devoted to this subject and we
are bound to repeat some well known facts. Our reason for this discussion is
that the IR/UV mixing , which will be central in the gravitational case, makes
an interesting appearance in a much better understood electric case. We will
see that there are unusual singularities in the in/ in propagators. Our
calculation here is very close to the one performed in the old paper
\cite{Nikishov} \cite{starob} (see also \cite{pol11}, \cite{gitman}) , but the
results seem new. The main goal of this section is to prepare for the
gravitational case.

Let us start with the standard set up ( see e.g. \cite{cooper} and references
therein) for the homogeneous , time dependent electric field $E(t)=\overset
{\cdot}{A}$, interacting with the complex scalar field $\varphi$ , which
satisfies the Klein- Gordon equation. We are interested in the case when the
field is turned on adiabatically, stays constant for a while , and turned off.
The field $\varphi$ has the standard mode expansion%
\begin{align}
\varphi & =\sum(a_{k}f_{k}^{\ast}+b_{k}^{+}f_{k}),\\
\overset{\cdot\cdot}{f}_{k}+\omega_{k}^{2}(t)f_{k}  & =0\nonumber
\end{align}
where $\omega_{k}(t)=\sqrt{(k-A(t))^{2}+k_{\perp}^{2}+m^{2}}.$ In the
adiabatic case we can use the WKB approximation for the modes. Let us assume
that $A(t)$ behaves qualitatively as a function $ET\tan(t/T)$ where $T$ is a
large adiabatic parameter. \ In this case we can fix the wave function to be a
single exponent in the past, $k\gg A(t).$ At the later time , when $k\simeq
A(t),$ the WKB approximation breaks down and the particles start to be
created. As the time goes to infinity, the WKB is working again, but the
reflected wave appears. So,%
\begin{align}
f_{k}  & \rightarrow\frac{1}{\sqrt{2\omega_{k}}}e^{is_{k}(t)},k\gg A(t)\\
f_{k}  & \rightarrow\frac{1}{\sqrt{2\omega_{k}}}(\alpha e^{is_{k}}+\beta
e^{-is_{k}}),k\ll A(t)
\end{align}
where $s_{k}(t)=\int_{0}^{t}\omega_{k}(t^{\prime})dt^{\prime}$ and
$\alpha,\beta$ are the Bogolyubov coefficients. These coefficients of course
depend on \ $k_{\perp}^{2}+m^{2}$ and, generally speaking, on $k$ . However,
it is important to realize that if \ $A(-\infty)\ll k\ll A(\infty),$we can
approximate $A(t)\simeq Et$ and in this case , by the transformation
$t\rightarrow t-k/E$ \ the $k$ can be removed from the wave equation. Thus
within this range the Bogolyubov coefficients do not depend on $k.$ Outside
this range they quickly vanish. These properties are the consequences of the
time - translation invariance of the constant electric field.

Now ,let us calculate various Green functions. Let us begin with
$F=\langle\varphi^{\ast}(z,t,x_{1\perp})\varphi(z,t,x_{2\perp})\rangle.$This
is the correlation between two points , separated in perpendicular direction,
but having coinciding projection on the direction of the electric field. Using
the above formulae we get
\begin{equation}
F=\int_{{}}dkdk_{\perp}e^{ik_{\perp}(x_{1}-x_{2})}|f_{k}(t)|^{2}=F^{(0)}+2\int
dk_{\perp}|\beta|^{2}e^{ik_{\perp}(x_{1}-x_{2})}\int_{A(-\infty)-A(t)}%
^{0}\frac{dp}{2\omega_{p}}%
\end{equation}
where $F^{(0)}$ is nonsingular correlator without electric field, $p=k-A(t),$
and the relation $|\alpha|^{2}-|\beta|^{2}=1$ has been used. The integral is
dominated by $p\gg m,k_{\perp},$ in which case $\omega_{p}\approx|p|,$and thus
we get the following contribution
\begin{equation}
F=\Phi(x_{1}-x_{2})\log(\frac{ET}{m})
\end{equation}
where $T$ is the IR\ cut off - the total time the electric field was on. Here
$\Phi(x)$ is a Fourier transform of $|\beta(k_{\perp})|^{2}$

If we look at the correlators with different $z$ we get%
\begin{equation}
F(z_{1}x_{1},z_{2}x_{2})=\Phi(x_{1}-x_{2})\log(\frac{1}{m|z_{1}-z_{2}|})
\end{equation}
provided that $\frac{1}{m}\gg$ $|z_{1}-z_{2}|\gg\frac{1}{ET}$. Such a
singularity at the non-coinciding points is quite unusual. Even more
interesting is the fact that the above formula is applicable only if
$ET<\Lambda_{UV}.,$where $\Lambda_{UV}$ is the ultraviolet cut-off. When
$ET\sim\Lambda_{UV}$ \ we have a remarkable \ situation - large time phenomena
depend on the ultraviolet completion of the theory, thus explicitly breaking
renormalizibility. When interactions are included, we encounter a problem (not
yet solved) of summation of leading logarithms.

It must be stressed that these interesting phenomena are lost if we use in/out
propagators so they describe \emph{non-equilibrium non-renormalizability
(NENR).. }It is easy to give a physical interpretation to the NENR. Imagine
two pairs created far away from each other. The electron from one pair moves
in the electric field towards the positron of another pair. When they collide,
their center of mass energy is of the order of $ET.$If $ET\sim\Lambda_{UV}$ ,
it is clear that the scattering amplitudes must depend on a particular
ultraviolet completion of the theory. At the same time, renormalizability is
just the statement of independence of the UV completion. It is not working in
our case, at least perturbatively. \ The appearance of the scale $T$
\ indicates the breakdown of the translation symmetry. Beyond the perturbation
theory it is possible that the collisions of created particles will establish
an equilibrium without the runaway phenomenon, described above. The outcome
depends on the high energy behavior of the cross-sections.

\section{The Gibbons - Hawking temperature and Jeans instability}

It may be worthwhile to discuss quantum instabilities of the de Sitter space
from a similar point of view. Let us explore to what extent one can use the
following naive ideas. Gibbons and Hawking \cite{gibbons}\ showed that the dS
space has intrinsic temperature defined by the area of the horizon, very much
like the Schwarzschild black hole. Does it make sense to say that as a result
the space simply evaporates ? . Also, in a thermal bath we would expect the
Jeans instability \cite{gross} . But our "bath" is unusual. The statement in
\cite{gibbons} was that any geodesically moving detector will measure the
above temperature. However already these authors noticed that there is
something funny about this statement. Unlike the usual heat bath, observers
moving relative to each other will not notice the Doppler shift. Also, this
temperature, unlike a physical heat bath, does not destroy the dS isometries.

To clarify the meaning of these notions we will try a different setting of the
problem, already touched upon in \cite{pol11} . Let us examine the space with
the warp factor $a(t)=\exp[T\tanh(\frac{t}{T})]$ . It starts with the
Minkowski space at $t\rightarrow-\infty$ ,then becomes the expanding dS space
for a long while ,and then returns to Minkowski space once again. When $T$ is
large, the quantum field theory on this background can be treated
semiclassically most of the time. We will concentrate on the particle
production in this situation. The important question is whether the particles,
produced in the future, will generate Jeans instability. Another related test
is to take the gravitational production of the electrically charged particles
and see if they can cause the Debye screeening and the plasma waves.
Incidentally, the same question can be posed in the case of electric
production . It has been addressed long ago in a nice paper \cite{cooper}
which, by the numerical simulations, showed the presence of the plasma
oscillation .

Let us find the Green functions near future infinity. The key point is that
for the large $T$ we can precisely match the plane waves at $\pm\infty$ with
the Bessel -like solutions of the Klein-Gordon equation in the dS space and
thus fully determine the reflection coefficient responsible for particle
creation. We have the Jost function $f_{k}(t)\rightarrow\frac{1}{\sqrt
{2\omega_{k}^{-}}}e^{i\omega_{k}^{-}t}$ as $t\rightarrow-\infty.$ Here
$\omega_{k}\left(  t\right)  =\sqrt{m^{2}+k^{2}/a^{2}(t)}$ and the $\pm$
\ superscripts refer to the limits $t\rightarrow\pm\infty$ \ \ \ . As we go to
the dS region, $|t|\ll T,$ this solution matches the Bunch-Davies mode
$h_{i\mu}(ke^{-t}).$ When we proceed to the "cash register" at future infinity
we find that the Bogolyubov coefficients
\begin{equation}
f_{k}(t)\rightarrow\frac{1}{\sqrt{2\omega_{k}^{+}}}(\alpha e^{i\omega^{+}%
t}+\beta e^{-i\omega^{+}t})
\end{equation}
come from the Hankel function and given by $a(\pm\mu).$ As we see they are
independent of $k$ (a consequence of \ dS symmetry). This is asymptotically
exact if
\begin{equation}
ma(-\infty)\ll k\ll ma(\infty)
\end{equation}
Outside this interval the particle creation quickly disappears. The Green
function at future infinity, derived from the above modes, is given by
\begin{equation}
\langle T\varphi(t)\varphi(t^{\prime})\rangle=\frac{1}{2\omega}[(1+|\beta
|^{2})e^{-i\omega|t-t^{\prime}|}+|\beta|^{2}e^{i\omega|t-t^{\prime}|}%
]+\frac{1}{2\omega}[\alpha\beta^{\ast}e^{-i\omega(t+t^{\prime})}+c.c]
\end{equation}
here $\omega=\sqrt{m^{2}+k^{2}/a^{2}(\infty)}$ . The coefficients $a(\pm\mu)$
\ satisfy the Wronskian relation $|a(\mu)|^{2}-|a(-\mu)|^{2}=1/2\mu$ as well
as \cite{pol11} $|a(-\mu)|^{2}=e^{-2\pi\mu}|a(\mu)|^{2}$ \ which leads to the
formula\footnote{The appearence of the Bose distribution from the Hankel modes
was also known to J. Maldacena.
\par
(private communication)}%
\begin{equation}
|\beta|^{2}=\frac{1}{e^{2\pi\mu}-1}%
\end{equation}
We can now interpret the above formula . The first term in ( ) is the thermal
propagator at the Gibbons- Hawking temperature ($1/2\pi$ in our units). At the
same time the full propagator is NOT thermal - it was obtained by the
evolution of the pure state. However , if we coarse- grain this pure
propagator, the second term goes to zero due to infinite oscillations and we
get the thermal state. It is also clear that the Doppler shift will be absent,
since as $a(\infty)\rightarrow\infty,$ the momentum dependence disappears from
the formulae.

We shall add to it a general comment concerning the purity of the propagator.
One might think that the thermal propagator can't be distinguished from the
one describing the $N$-th excited state. Indeed they have the same form except
in the thermal state $N$ is given by the Bose distribution. However there is a
major difference when we consider the four point function- there is the Wick
theorem in thermal state and no such relation in the excited state.

Debye screening means that the photon acquired \ a mass, while the Jeans
instability is an imaginary mass of the graviton. Formally they both appear
because the continuity equation for the polarization operator doesn't force it
(except in the vacuum) to be zero at zero momentum \cite{gross}. This happens
because the one loop polarization operator has singularities at zero energy
and momentum. If the particles in the loop have the energy $E_{1}$ and $E_{2}$
, the polarization operator contains a singularity at the frequency
$\omega=E_{1}\pm E_{2}$ . The plus sign describes pair creation by the
external quantum, while the minus term comes from the collision of the
external quantum with the preexisting particle. As a result , when
$\omega,q\rightarrow0$ we have a non-commutative limits, first noticed by A.
B. Migdal in the theory of Fermi liquid. \ For the non-interacting particles
the polarization operator has the form%
\[
\Pi(q,\omega)\symbol{126}\int\frac{dp}{\omega_{p}\omega_{p+q}}\frac
{[N(\omega_{p+q})-N(\omega_{p})](\omega_{p}+\omega_{p+q})}{\omega^{2}%
-(\omega_{p}-\omega_{p+q})^{2}}+...
\]
where by dots we denoted the non-singular part and $N$ is the Bose
distribution. This term creates real mass for photons ( plasma waves ) and
imaginary mass for gravitons ( instability).

In the present case the question of the Jeans instability requires further
study. The particle distribution $N(q)=|\beta|^{2}$ is given by () and is
thermal only for $q\ll ma(\infty)$ \ while a simple computation of the
polarization operator includes the values of $\ q\sim ma(\infty)$ for which
the distribution is $q$ dependent \ and noneqilibrium . Most likely, the
interaction between the particles will equilibrate this distribution at some
non-universal temperature and produce the Jeans instability. But this remains
to be calculated. It is interesting to notice that in the purely electric case
numerical simulations in \cite{cooper} led to the plasma oscillations.

In the next section we will discuss a different type of instability, related
to particles interaction.

\section{Unstable symmetry of the Poincare patch. A conjecture.}

The Poincare patch is a one half of the global de Sitter space, representing
expanding or contracting universes. It has some fatal attraction for
cosmological theories, from the steady state universe to the inflationary one.
It seems well fitted for the observed cosmological expansion. However, in
quantum theory one must be careful with the geodesically incomplete spaces,
since particles may inadvertently disappear, making havoc of unitarity.

In this section we will show that the standard Bunch - Davies vacuum in the
Poincare patch preserves ( in a non-trivial way) the full de Sitter symmetry.
At the same time any small variation of this state leads to a complete
destruction of symmetry, due to the violent particle production. \ 

The dS space is described by the hyperbolic unit vector $n$ , satisfying
\ $n_{+}n_{-}+n_{\perp}^{2}=1$ with $n_{\pm}=n_{1}\pm n_{0}$ . The Poincare
patch is defined by $n_{+}\geqslant0$ . When we consider a field theory in
this space, we must integrate over the position of the vertices in Feynman's
diagrams. In the full space the measure of integration is just ($dn)\delta
(n_{+}n_{-}+n_{\perp}^{2}-1).$ In the Poincare patch this measure must be
multiplied by $\vartheta(n_{+}).$ This factor, generally speaking, breaks the
dS symmetry. Indeed, consider a rotation $\delta n_{+}=\omega_{+\perp}n\perp$
\ It obviously changes the measure, since the variation of the step function
gives the factor $n_{\perp}\delta(n_{+}).$ Let us study the effect of this
variation. Consider a vertex located at the point $n$ in some generic
Schwinger - Keldysh diagram. It is attached to the rest of the diagram by a
bunch of the Green functions , $\prod\limits_{A}G(n_{A}n).$ One should also
remember that each point of the diagram carries $+$ or $-$ \ signs. The change
of the amplitude $F$ is given by%
\begin{equation}
\delta F\sim\sum_{\pm}\int dn_{\perp}dn_{-}n_{\perp}\prod G(n_{A}n)\mid
_{n_{+}=0}%
\end{equation}
We have to analyze this contribution for the various $\pm$ combinations. \ For
that we have to remember that in the Bunch Davies vacuum all Schwinger -
Keldysh (SK) functions are various boundary value of the single analytic
function, which depends on $z=(nn^{\prime}).$ One also should remember that
the ordering of the operators ( the arrow of time ) could be performed with
the respect to $n_{+}$ or with the respect to $\ n_{0}$ . These two are
equivalent. Indeed, the relation $n_{0}^{\prime}\leq n_{0}$ doesn't
automatically imply that $n_{+_{{}}}^{\prime}\leq n_{+}$ but if the latter is
not true, the interval between the two points is space-like and the order of
operators is irrelevant. To check this we write $\ 2n_{0}=n_{+}+n_{-}$ and
notice that the interval $s=(n_{+}^{\prime}-n_{+})(n_{-}^{\prime}%
-n_{-})+(n^{\prime}-n)_{\perp}^{2}\geq0$ , if $n_{0}^{\prime}\leq n_{0}$ but
$n_{+}^{\prime}\geq n_{+}$ .

The SK set of the Green functions is given by%
\begin{align*}
G_{++}(n,n^{\prime})  & =g(nn^{\prime}-i0)\\
G_{+-}(n,n^{\prime})  & =g(nn^{\prime}-i\epsilon sgn(n_{+}^{\prime}-n_{+}))\\
G_{-+}(n,n^{\prime})  & =G_{+-}^{\ast}\\
G_{--}(n,n^{\prime})  & =G_{++}^{\ast}%
\end{align*}
where the analytic function $g(z)$ in the case of the BD vacuum has a single
branch point at $z=1,$ \ and is real for $z\leq1$ ( the space-like
separations). Substituting these functions into the above formula and taking
into account that $n_{+}=0$ is the ultimate past, we get%
\begin{equation}
\delta F\sim\operatorname{Im}\int dn_{\perp}n_{\perp}\int dn_{-}\prod
g(n_{A+}n_{-}+n_{A\perp}n_{\perp}-i0)
\end{equation}
From this we see that we can close the contour of the $n_{-}$ integration in
the lower half-plane (since $n_{A+}\geq0$ ) and the integral, in the case of
the BD vacuum, is zero.

However, this symmetry is likely to be unstable. If we change the B.D. Green
functions in the past, in such a way that the dS symmetry is preserved on the
classical level, the quantum effects will unravel this symmetry. Indeed, let
us consider the state with a non-zero occupation numbers. The simplest
example, describing the constant occupation numbers, leads to the Green
function%
\begin{equation}
G^{(N)}(z)=(1+N)g(z-i0)+Ng(z+i0)
\end{equation}

This function \ is de Sitter symmetric. Notice, however, that it is not one of
the Mottola- \ Allen $\alpha-$ vacua. The latter describe the Fock vacua,
while in our case we average with respect to the excited state. One popular
"objection" against such states ( and against $\alpha$ -vacua) is that in the
short distance limit these Green functions don't reduce to the Minkowsky
vacuum Green function. This objection is unjustified. In the geodesically
incomplete spaces the Green functions are determined by the choices made
behind the horizon. As an analogy, consider the Rindler space. Field theory in
this space has a well defined vacuum. However, if we assume that the full
space is in the ground state, the induced Green function in the Rindler wedge
is the thermal one.

Returning to the case above, we can't close the contour and the dS symmetry is
badly broken. Moreover, as we will see below, the breaking is amplified by the
logarithmic IR divergence. In this sense the Poincare patch symmetry is likely
to be unstable, at least in the linear approximation. We \ will consider now
the general variation of particle numbers. To simplify notations, we will
discuss a $\varphi^{3}$ theory.

In the previous papers \cite{pol11} , \cite{pol10} we introduced various
production amplitudes. They describe the following allowed decays and fusions.
We can have the process $vac\Rightarrow(q)+(k)+(-k-q).$with the amplitude
$A(q,k)$ and the inverse process with the amplitude $A^{\ast}$ ( the notations
mean that the vacuum decays onto first excited states with the momenta
$q,k,-k-q$ ). We can also have the particle decay, $(q)\Rightarrow(k)+(q-k)$
with the amplitude $B(q,k)$ , and most importantly, the bremsstrahlung
radiation, $(k)\Rightarrow(k-q)+(q).$ with the amplitude $C(q,k).$ Using the
standard mode expansion, one expresses these amplitudes in terms of overlap
integrals \cite{pol10}%
\begin{align}
A(q,k)  & =\int_{\tau}^{\infty}d\tau\tau^{\frac{d}{2}-1}h(q\tau)h(k\tau
)h(|k+q|\tau)\\
B(q,k)  & =\int_{\tau}^{\infty}d\tau\tau^{\frac{d}{2}-1}h^{\ast}(q\tau
)h(k\tau)h(|q-k|\tau)\\
C(q,k)  & =\int_{\tau}^{\infty}d\tau\tau^{\frac{d}{2}-1}h(q\tau)h^{\ast}%
(k\tau)h(|k+q|\tau)
\end{align}
where $h(x)$ are proportional to the Hankel functions $H_{i\mu}^{(1)}(x)$ .
The self-energy corrections to the Schwinger - Keldysh propagators are
expressed in terms of the products of these amplitudes, for example the +/-
correction to the two point function $\langle\varphi(q\tau)\varphi
(-q\tau)\rangle$ is proportional to the integral $\int d^{d}k|A(q,k)|^{2}$
(see \cite{pol11} for the details). The infrared corrections ( not
divergences) appear because for $k\gg q$ , $|A(q,k)|^{2}\sim k^{-d}.$ , which
generates a logarithmic correction $\log(\frac{1}{q\tau}).$ These corrections,
in agreement with the above theorem, do not violate the de Sitter symmetry.
Another reason for the absence of \ the divergence in this case follows from
the observation \cite{marol10} that the one point function in the Poincare
patch is the same as in the static patch. However, in the latter case we have
thermal equilibrium and the collision integral is equal to zero. This is also
in agreement with the absence of corrections to the self-energy part
\cite{leblond}

The situation changes drastically if we start with the non-zero occupation
numbers . Consider a bare Green function%
\begin{equation}
\langle\varphi(q\tau)\varphi(-q\tau^{\prime})\rangle=(1+n(q))h^{\ast}%
(q\tau_{>})h(q\tau_{<})+n(q)h^{\ast}(q\tau_{<})h(q\tau_{>})
\end{equation}
which corresponds to the averaging with the respect to the excited states with
the occupation numbers $n(q).$The major difference comes from the $C$ -
amplitude (which doesn't appear when $\ n(q)=0)$ . As is seen from the above
formulae, the integrand for this amplitude is not oscillating at very large
$\ k$ and this changes the asymptotic behavior of $\ C$ . As $k\rightarrow
\infty$ we have%
\begin{equation}
C\simeq\frac{1}{|k|\frac{{}}{{}}}\int d\tau\tau^{\frac{d}{2}-2}h(q\tau
)e^{iq\cos\theta\tau}%
\end{equation}
where $\theta$ is the angle between $k$ and $\ q$ . We can now evaluate the
contribution of the $\ C$ - process to the occupation numbers. Keeping only
the linear term in $n(k)$ in the collision integral, we obtain%
\[
\Delta n(q)\propto\int d^{d}k|C(q,k)|^{2}n(k)\propto\frac{1}{q^{d-2}}\int
\frac{d^{d}k}{k^{2}}n(k)
\]
This integral can become divergent for the perturbations concentrated at large
$\ k.$ This is an indication of the linearized instability of the Poincare
patch. \ Indeed, even if we assume that the original perturbation $n(k)$ is
vanishing so fast that the integral converges, in the second iteration one
gets $\Delta n(k)\sim1/k^{d-2}$ which leads to the IR divergence. One needs to
study the collision integral in full to make a definite conclusion beyond the
linear approximation.. We will not proceed with it here, since our main
interest is in the geodesically complete case. Notice, however, that the
contribution of $A$ and $B$ are suppressed by the factor $(q/k)^{d-2}$ .

Let us stress the difference between the dS and the Minkowski space. In the
latter case we can also change the occupation numbers, breaking the symmetry
at the initial moment. The system then will tend to thermal equilibrium. The
temperature for the small initial perturbations will remain small. In the dS
case the conserved energy is not positive definite, there is no H-theorem, and
we may experience the instability.

\section{Hyperbolic motion, an analogy}

The picture described above is analogous to the classic problem of radiation
of the accelerated charge. We will briefly describe it to stress the analogy.
The probability of radiation is given by the "golden rule"%
\begin{equation}
w\sim\int|J_{\mu}(k)|^{2}\theta(k_{0})\delta(k^{2})d^{4}k
\end{equation}
where the current \ (or vertex operator) is given by \ $J_{\mu}(k)=\int
dsx_{\mu}^{\prime}(s)e^{ikx(s)}.$ , with $x(s)$ being a trajectory of the
charge. It is instructive to perform the k- integration and obtain the
expression for the probability
\begin{equation}
w\sim\int\int\frac{ds_{1}ds_{2}(x^{\prime}(s_{1})x^{\prime}(s_{2}))}%
{(x(s_{1})-x(s_{2})+i\epsilon)^{2}}%
\end{equation}
and for the radiated energy (obtained by adding $k_{\mu}$ factor in (19 ))%
\begin{equation}
P_{\mu}\sim\int\int\frac{ds_{1}ds_{2}(x^{\prime}(s_{1})x^{\prime}%
(s_{2})(x_{\mu}(s_{1})-x_{\mu}(s_{2}))}{(x(s_{1})-x(s_{2})+i\epsilon)^{4}}%
\end{equation}
Here $\epsilon$ is an infinitesimal time-like vector. The first expression is
an interesting version of the Wilson loop ( different from the standard one by
the $i\epsilon$ prescription). This prescription makes (21) non-zero. Using
the antisymmetry of this expression we get the standard Lorentz-Dirac formula
\begin{align}
P_{\mu}  & \sim\int ds[v_{\mu}^{\prime\prime}-(vv^{\prime\prime})v_{\mu}]\\
(v  & =x^{\prime},v^{2}=1)
\end{align}
The condition of the constant rate can be formulated as a symmetry of the
trajectory. In the case of the inertial motion of the charge , the symmetry is
simply $x(s+t)=x(s)+c$ and we have no radiation. In the hyperbolic case the we
have
\begin{equation}
x_{\mu}(s+t)=\Lambda_{\mu\nu}(t)x_{\nu}(s)
\end{equation}
where $\Lambda$ is a Lorentz transformation. This is the analogue of the dS
symmetry. The Lorentz -Dirac equation has the form%

\begin{equation}
v_{\mu}^{\prime}=\varepsilon_{\mu\nu}v_{\nu}+\frac{2\alpha}{3}[v_{\mu}%
^{\prime\prime}-(vv^{\prime\prime})v_{\mu}]
\end{equation}
and the last term vanishes for the hyperbolic motion. This is analogous to the
preservation of the dS symmetry in the preceeding section. It means that there
is no radiation from the hyperbolic motion. However this solution is unstable
. It is well known that the equation has runaway solutions violating the above
symmetry. Under a small perturbation \ the particle begins to radiate. Dirac
tried to impose initial and final conditions together in order to avoid the
runaway. In our language this amounts to considering in/out matrix elements.
As we mention above, there is no IR divergences in these amplitudes.

\section{Upside -down description of the global dS space}

\bigskip We shall now apply the above methods to the global ( geodesically
complete ) space. This space is known to be classically stable \cite{anderson}
in the sense that the Cauchy problems at past infinity are well posed. In a
vulgar language of a field- theorist, this means that the tree S-K diagrams
(representing a perturbative solution of the Cauchy problem ) are not IR
divergent. However, this is not true for quantum corrections.

Once again we will consider an interacting massive field on the dS background
and assume that the interaction is adiabatically turned on near the past
infinity. If we are interested in the one point functions, we can achieve a
considerable simplification by the following trick. Suppose that the point in
consideration is given by a hyperbolic vector $n^{\ast}$, which we can always
chose to have $\ n_{-}^{\ast}\geq0.$ In the Schwinger -Keldysh approach all
interactions take place inside the past light cone, defined by $\ (n-n^{\ast
})^{2}\leq0$ and $n_{0}\leq n_{0}^{\ast}$ . Rewriting this as $\ (n_{-}%
-n_{-}^{\ast})(n_{+}-n_{+}^{\ast})\leq0$ \ and $n_{+}-n_{-}\leq n_{+}^{\ast
}-n_{-}^{\ast}$ \ we conclude that $n_{-}\equiv1/\tau\geq0$ . Therefore, as
far as one point functions are concerned, the global dS space is identical to
the Poincare patch, turned upside down, with $\tau=0$ being past infinity and
the time development defined by the increasing $\tau$ ( which we may call the
"Anti- Poincare patch)

This statement may looks strange, since we ended up with the contracting
patch, while the global dS contains both contraction and expansion. It should
be remembered, however, that we are dealing here with the one-point functions
and it is impossible to distinguish contraction from expansion while staying
at one point. \ Moreover, the standard definition of the red shift \ refers to
two static observers ( to eliminate the Doppler effect from the peculiar
velocities). But in the case of the dS space the "static $^{{}}$observer" in
the Poincare coordinates is generally non-static in the global coordinates.

The global two point function can also be described in the above patch,
provided that the two points are lying inside the horizon. It is easy to see
that this restricts the positions so that $(n_{1}n_{2})\geq-1.$ As we cross
this boundary , the Green function can be non-analytic. Yet another
interesting feature of the Anti- Poincare patch is that the derivation of the
dS symmetry given in the preceeding section is not applicable. Technically
this happens because we reverse the arrow of time. This leads to a different
$i$ $\epsilon$ prescription in the formula ( ) and this integral isn't zero anymore.

Let us begin with the processes described by the $A$ - amplitude. In the Anti
- Poincare case it is given by the same expression as in the usual case,
except that the integral over $\tau$ is going from zero ( past infinity ) to
the present moment. We also have to introduce the IR -cut-off $\varepsilon$,
defined as the time in the past when the interaction was turned on. We have
\begin{equation}
A(q,k)=\int_{\varepsilon}^{\tau}\frac{d\tau_{1}}{\tau_{1}}\tau_{1}%
^{d/2}h(q\tau_{1})h(k\tau_{1})h(|k-q|\tau_{1})
\end{equation}
and analogous formulae for $\ B$ and $C$ amplitudes. The dominant contribution
comes from $\ k\gg q.$ Just as in \cite{pol11} , using the expansion at
$\ x\rightarrow0,\ h(x)=a(\mu)x^{i\mu}+a(-\mu)x^{-i\mu}$ \ for the Hankel's
functions , normalized by $h(x)\rightarrow(2x)^{-\frac{1}{2}}e^{ix}$ as
$x\rightarrow\infty$ , we obtain%
\begin{align}
A(q,k)  & \rightarrow k^{-\frac{d}{2}}[a(\mu)g(\mu)(\frac{q}{k})^{i\mu}%
+(\mu\rightarrow-\mu)]\\
g(\mu)  & =\int_{0}^{\infty}dxx^{d/2-1+i\mu}h^{2}(x)
\end{align}
\qquad This asymptotics assumes that
\begin{equation}
1/\varepsilon\gg k\gg\max(q,1/\tau)
\end{equation}
since under these conditions we can drop the limits in the above integral. The
contribution to the Green function can be calculated using the Schwinger -
Keldysh \ approach and in this order we have%
\begin{align}
G_{++}^{(1)}(q,\tau)  & =G_{++}\Sigma_{++}G_{++}-G_{+-}\Sigma_{-+}%
G_{++}+(+\leftrightarrow-)\nonumber\\
G_{++}(q,\tau_{1},\tau_{2})  & =(\tau_{1}\tau_{2})^{d/2}h^{\ast}(q\tau
_{>})h(q\tau_{<})
\end{align}
Combining these formulae, we get the general structure of the correction to
the Green function (for simplicity we look at the coinciding times )%
\begin{equation}
G^{(1)}=2N(q,\tau)|h(q\tau)|^{2}+\operatorname{Re}(\alpha^{\ast}(q,\tau
)h^{2}(q\tau))
\end{equation}
The same structure appeared in \cite{pol11} , however in the global dS space
it has very different meaning. Let us look at the $\Sigma_{+-}$ term which
contains a square of the $A$ amplitude. It gives a contribution to $N$
\begin{equation}
N(q,\tau)\sim\int|A(q,k)|^{2}d^{d}k
\end{equation}
\qquad In the Poincare patch it gave a contribution of the order of
$\log(1/q\tau)$ which was important only if $q\tau\ll1.$ These logarithms
modify the large distance behavior of the Green function but do not contain
the infrared \emph{divergence, }responsible for the secular change of the
vacuum ( and for the irreversible time in the Boltzmann equation).\emph{\ }%
\ This fact is also a consequence of the above theorem on the symmetry of the
Poincare patch. It all changes in the global case. Using ( ) we have
\begin{equation}
|A(q,k)|^{2}=(|ag|^{2}+|\widetilde{a}\widetilde{g|}^{2})k^{-d}+...
\end{equation}
where tilde means the change $\mu\rightarrow-\mu$ and the dots stand for the
oscillating terms. As a result we get \
\begin{equation}
N(q,\tau)\propto(|ag|^{2}+|\widetilde{a}\widetilde{g}|^{2})\log(\frac{\tau
}{\varepsilon})
\end{equation}
in the region $q\tau\sim1.$This logarithm is equal to the proper time passed
since the interaction was switched on, up to the present moment and thus we
have a linear divergence in time, just as we have one in the Boltzmann
equation when the collision integral is non-zero ( we mean the kinetic
equation $\frac{dn}{dt}=C[n],$ being iterated with the non-equlibrium initial
conditions). \ The above result comes from the $\Sigma_{+-}$ term. The terms
with $\Sigma_{++}$ and $\Sigma_{--}$ generate the contributions proportional
to $h^{2}(q\tau)$ and its complex conjugate. As was explained in \cite{pol11},
they contain a product of $\ A$ and $B$ amplitudes,leading to
\begin{equation}
\alpha(q,\tau)\sim\int d^{d}kA(q,k)B^{\ast}(q,k)\sim a\widetilde{a}%
(|g|^{2}+|\widetilde{g}|^{2})\log(\frac{\tau}{\varepsilon})
\end{equation}
The $C$ amplitude doesn't appear in this order, since we started from the
vacuum state and the $\ C$ - process requires non-zero occupation numbers.

It is easy now to evaluate the one point function in this approximation ( see
also \cite{pol11} ). We have%
\begin{equation}
\langle\varphi^{2}(n)\rangle\sim\tau^{d}\int d^{d}q|h(q\tau)|^{2}\int
d^{d}k|A(q,k)|^{2}%
\end{equation}
The integral is divergent at large physical momentum $p=q\tau.$It is cut off
in two different ways. First, the physical momentum must satisfy $p\ll
\Lambda_{UV}$ , where $\Lambda_{UV}$ is the ordinary ultraviolet cut-off.
Notice, that without interaction the second integral in this formula is absent
and we have the standard flat space result $\langle\varphi^{2}(n)\rangle
^{(0)}\sim(\Lambda_{UV})^{d-1},$ since $|h(p)|^{2}\sim1/p$ \ as
\ $p\rightarrow\infty$ . This is just an \ \ uninteresting UV correction to
the cosmological constant. \ 

Let us evaluate the above expression. For that we rewrite the formula (17)
taking into account the domain of the integration. We have%
\begin{equation}
g(\mu,\tau,\varepsilon)=\int_{k\varepsilon}^{k\tau}dxx^{d/2-1+i\mu}h^{2}(x)
\end{equation}
We see that the maximal contribution to the one point function comes from the
region%
\begin{equation}
m\ll q\tau=p\ll k\tau\ll\min(m\frac{\tau}{\varepsilon},\Lambda_{UV})
\end{equation}
here we introduced explicitly the mass $\ m$ of our scalar field, which was a
dimensionless number before ( measured in the units of the Hubble constant).
We see the same non-commutativity of limits as for the electric case, namely,
if $\Lambda_{UV}\gg m\frac{\tau}{\varepsilon}$ we get%
\begin{equation}
\langle\varphi^{2}(n)\rangle^{(1)}\sim m^{d-1}(\frac{\tau}{\varepsilon}%
)^{d-1}+O(\log(\frac{\tau}{\varepsilon}))
\end{equation}
where the second term represents the contributions of the $\ B-$ amplitude,
proportional to $h^{2}(p)$ and its complex conjugate. These functions
oscillate at the large $p$ and thus do not produce the power UV divergence. \ 

A natural interpretation of the above formulae is the large blue shift of the
virtual particles. However, one must be careful with such an interpretation,
since , as was already mentioned , this notion is coordinate -dependant due to
the Doppler shift. Although on a technical level we are working with the
universe contracting to zero and having large blue shifts, we apply these
results to describe a non-singular global dS space.

A question, arising at this stage, is the summation of the leading logarithms,
the contributions of the order $(\lambda\log(\frac{\tau}{\varepsilon}))^{n}$
where $\lambda$ is a coupling constant. The answers to this question is very
different in the cases of the Poincare patch and the global AdS. In both cases
we begin in the lowest order with the contributions to the self-energy part
(written schematically) $\Sigma(q)\sim\int G(k)G(q-k)d^{d}kd\tau_{1}d\tau_{2}$
. The lowest logarithmic contribution, discussed above, is coming from the
domain $k\sim1/\tau_{1,2}\gg q.$ The next correction comes from the correction
to $G$ in the above integral , $G^{(1)}(k)\propto\int G^{2}(k^{\prime}%
)d^{d}k^{\prime}d\tau_{3}d\tau_{4}.$

In the Poincare patch this correction is negligible, since as we indicated
before, the logarithms in $G(k)$ arise only when $k\tau\ll1.$ As a result, in
this case higher order corrections to the self-energy part are not logarithmic
since they involve the Green functions at the momenta $k\sim1/\tau.$ This
conclusion agrees with the work \cite{leblond}.

In the global AdS we found that the logarithmic contribution to the Green
function $G(q,\tau)$ is present for any $q\tau$ , provided that $q\varepsilon
\ll1$ and $\frac{\tau}{\varepsilon}\gg1$ . Hence, in this case the self-
energy part receives non-trivial corrections.

Qualitatively they can be viewed as a cascade production and absorption of
particles with the momenta $q\ll k\ll k^{\prime}\ll...$ . With each generation
the momentum increases. This is very similar to the \textit{direct cascade} in
the theory of turbulence .

\section{An even-odd curiosity and four-valued time .}

It was already noticed \cite{pol10} that in the lowest order of perturbation
theory the amplitudes for particle creation are zero. This has an interesting
consequence for the infrared divergences. Let us repeat the calculations of
the previous section by keeping explicitly the contracting phase \cite{pol11}
and using the global eigenmodes. The Schwinger - Keldysh rules are modified,
making the flow time four- valued ( forward / backward and contracting /
expanding ) \ As was explained in \cite{pol11} the logarithmic corrections
arise in the \ region where the global modes, which are the Legendre functions
can be approximated by the Hankel \ \ functions as following%
\begin{equation}
f_{q}(t)\rightarrow\{%
\begin{array}
[c]{ccc}%
\tau^{d/2}h(q\tau) &  & \\
(-\tau^{\prime})^{d/2^{{}}}h(-q\tau^{\prime}) &  &
\end{array}
\end{equation}
Here $\tau=e^{-t}$ \ and $t\rightarrow+\infty$ in the upper expression. This
term describes the expanding stage. In the lower expression $\tau^{\prime
}=e^{t}$ \ and $\ t\rightarrow$ $-\infty$ and thus describes the contracting
phase. In the Feynman rules for the interacting theory each vertex contains
not only the usual Schwinger - Keldysh $\pm$\ \ but also a summation over
expanding/ contracting locations, making time a four-valued quantity.

In the present case the contributions to the $A$ amplitude come from the both
stages. \ We have the total contribution in the form%
\begin{equation}
A=A^{(c)}+(-1)^{d+1}A^{(e)}%
\end{equation}
\qquad\qquad\ where the amplitudes are given by the overlap integral (14 )
with the functions (40 ). \ Namely
\begin{align}
A^{(e)}  & =\int_{\tau}^{\infty}\frac{d\tau_{1}}{\tau_{1}^{d+1}}\tau
_{1}^{3d/2}hhh\\
A^{(c)}  & =\int_{-\infty}^{-\varepsilon}\frac{d\tau_{1}}{\tau_{1}^{d+1}}%
\tau_{1}^{3d/2}hhh
\end{align}
The sign factor in (41 ) comes from the $(-$ $\tau)^{d+1}$ in the phase
volume. When the dimension $D=d+1$ is even, there is a destructive
interference between contraction and expansion. This is seen from the identity%
\begin{equation}
\int_{-\infty}^{\infty}\frac{d\tau}{\tau^{d+1}}\tau^{3d/2}hhh=0
\end{equation}
since we can close the contour in the upper half-plane. \ In the previous
section we obtained the scale invariant behavior of the amplitudes in the
domain where the lower limits of the integral (42 ) can be set to zero. If we
can set to zero the upper limit of (43 ) also, we will get the cancellation in
the even dimension and doubling in the odd. \ The condition for this to happen
is to have $\tau_{1}\sim1/k\gg\tau$ in the first integral and $\tau_{1}%
\sim1/k\gg\varepsilon$ in the second. As before, we also need $k\gg q.$ With
these conditions we obtain the first correction to the occupation numbers in
the form%
\[
N(q,\tau\grave{)}\propto\log(\frac{1}{q\varepsilon})+(-1)^{d+1}\log(\frac
{1}{q\tau})\propto\{%
\begin{array}
[c]{c}%
\log(\frac{\tau}{\varepsilon})\\
\log(\frac{1}{q^{2}\tau\varepsilon})
\end{array}
\]
for even and odd cases. So, in the even case the instability is weaker than in
the odd. It is interesting that the above formula is different from (34 ) when
$q\tau\ll1.$ This is not surprising, since the equivalence of the anti -
Poincare and the global case works only for two point functions with the
points inside the light cone. The above condition violates this constraint.\ 

The above considerations refer to the lowest non-trivial order of perturbation
theory. It is not clear at this point if they survive in the higher orders.

\section{The Ansatz for the Green functions.}

In order to sum the leading logarithms we need an ansatz for the Green
functions. This problem has not been solved yet. In this section we will
indicate a plausible approach to its solution

\ . Let us first remember how the secular terms in the Green functions are
summed up in the case of the ordinary Boltzmann equation. In this case we have
two different time scales, one given by the characteristic energies of the
particles and another, much longer time scale - that of the free flight. As a
result, in the leading approximation, the Green functions retain their non-
interacting form but the occupation numbers, defining these functions are
slowly changing . The evolution of these numbers is governed by the collision
integral and as noticed in \cite{Polyakov} can be viewed as an IR analogue of
the renormalization group. So, the key point concerning kinetics is that the
weak enough interaction modifies the Green functions only through the slow
variation of the occupation numbers. Another point is that the kinetic
equation is just the Dyson equation \ in which the lowest order self-energy
part is expressed in terms of the dressed Green functions, while the vertex
corrections can be neglected.

Something very similar is also true in our case . The formula (20) shows that
the interaction \ leaves the Green function \ \ free, except that we have a
slowly varying factors $\ N(q,\tau)$ and $\alpha(q,\tau),$ the two soft modes.
These factors can be interpreted as the occupation. number and the
infinitesimal Bogolyubov rotation. Therefore we shall conjecture the following
structure\footnote{A similar ansatz was considered in \cite{akhmedov} and by
D. Krotov (unpublished) in the Poincare patch.} of the Green functions in the
global dS space%
\begin{equation}
G(q,\tau)=(1+N(l))f^{\ast}(q,\tau_{>})f(q,\tau_{<})+N(l)f^{\ast}(q,\tau
_{<})f(q,\tau_{>})
\end{equation}
where $l=\log\frac{\tau}{\varepsilon}$ and
\begin{align}
f(q,\tau)  & =\alpha(l)h(q\tau)+\beta(l)h^{\ast}(q\tau)\\
|\alpha|^{2}-|\beta|^{2}  & =1
\end{align}
This ansatz assumes that beyond the first approximation we get a finite
Bogolyubov transformation.\ When we substitute this expression into the
Schwinger- Keldysh self energy part, $\Sigma\sim\int GG$ we will get the
collision integral. It contains many terms involving products of
$N,\alpha,\beta.$ The overlap integrals $A,B,C$ (14 - 16 ) now contain the
functions $\ f$ instead of $h.$ The effect of the non-zero occupation numbers
, say, in the $\ A$ amplitude  is seen from the replacement of (32 ) by%
\begin{equation}
N(q,\tau)\sim\int d^{d}k|A(q,k)|^{2}%
\{(1+N(q)(1+N(k))(1+N(q-k))-N(q)N(k)N(q-k)\}+...
\end{equation}
The calculations in the preceeding sections indicate that these quantities
(with the logarithmic accuracy ) depend on its arguments in a specific way.
Namely, if we introduce the physical momentum $p=q\tau$ , the occupation
numbers at the fixed $p$ depend on $\tau$ only and not on $p\symbol{126}m$.
More precisely, there are extra logarithmic terms if $p\ll m$ \ , but , at
least in the leading order this domain is negligible. After going to the
scaling limit the above term becomes%
\begin{equation}
N(\tau)=\int_{\varepsilon}^{\tau}\frac{d\tau_{1}}{\tau_{1}}|A(\tau_{1}%
)|^{2}\{(1+N(\tau_{1}))^{2}(1+N(\tau))-N^{2}(\tau_{1})N(\tau)\}+...
\end{equation}
here the $\ A$ depends on $\tau$ through the Bogolyubov coefficients ; the
dots stand for the contribution of the $B$ and $C$ processes described above.
We need another equation for these coefficients. It follows from the matching
of the terms in the Green function proportional to $\ h^{2}(q\tau)$ and its
complex conjugate. This is a straightforward procedure, but it results in long
clumsy equations. Perhaps going to the \ geometrical variables of SL(2,C),
describing the Bogolyubov transformation, will bring some simplification.

There is, however, another problem with our ansatz. The $\ C$ - type
amplitudes, according to (14 ) contain the integrals involving the
combinations $\ hh^{\ast}$ which don't oscillate at very large $k$ . This can
cause a divergence which will the invalidate our scaling limit. The problem
has not been fully resolved. Let us notice that it is related to the
bremsstrahlung radiation of the soft quanta, moreover, the domain of large $k
$ corresponds to the flat space limit. The emission of such quanta produces
the well-known divergences in perturbation theory, which, however, disappear
under the appropriate treatment. We hope that the same cancellation occurs in
our case and the above divergences can be renormalized away. Formally, it is
easy to give a prescription which eliminates the power - like divergences. It
is sufficient to define the overlap integrals as an analytic continuation in
$d$ \ from the region $\ d<2 $ where they converge.

\section{Conclusions}

The main conclusion of this paper is that the de Sitter spaces of various
kinds are unstable on the quantum level. In this paper we discussed only the
interacting massive scalar fields , which is technically the easiest case.
\ It is of course very important to include massless gravitons in our picture.
Formally the massless case corresponds to taking $\mu=i\frac{d}{2}$ . The
overlap integrals become superficially divergent. However, this leading
divergence must cancel due to gauge invariance. We will be left with the
subleading divergence, similar to the ones considered above. A lot of work
remains to be done, mostly related to the proper treatment of gauge degrees of freedom.

The summation of the leading logarithms is another unfinished business. It is
likely that it will lead to the Landau poles in various observables and will
require new methods. Whatever the answer will be, this is a well posed
fascinating problem.

Another interesting problem is to consider instead the global dS space the
"centaur" - a Coleman - de Lucia bubble appeared through tunneling. It is
described by the same Feynman diagrams , but, as usual, the time variable is
imaginary under the barrier. It seems likely (but not proved) that dS symmetry
is unstable also in this case.

It may be too early to try to build models based on the above mechanisms (see,
however, \cite{klink} ) . Still it might be relevant in various cosmological
scenarios. Indeed, the standard inflation is based on the asumption of slow
roll towards zero cosmological constant. Even if we assume that the
cosmological constant problem is somehow solved, the question remains, why the
theory was originally in a highly excited state and not simply in the
Minkowsky vacuum. In the above picture the presence of the original
cosmological constant will make the steady state impossible.

I would like to thank E. Akhmedov, M. Anderson, D. Krotov, J. Maldacena and
especially V. Mukhanov for many useful discussions. This work was partially
supported by the NSF grant PHY-0756966

\end{document}